\documentclass[aps, pra, superscriptaddress, twocolumn, 10pt]{revtex4-1}

\usepackage{amsfonts,amssymb,amsmath,amsthm,longtable,array,tabularx}
\usepackage{graphics,graphicx,color}
\usepackage{epstopdf,epsfig}

\def\identity{\leavevmode\hbox{\small1\kern-3.8pt\normalsize1}}
\def\openone{\leavevmode\hbox{\small1 \normalsize \kern-.64em1}}
\newcommand{\be}{\begin{eqnarray}}
\newcommand{\ee}{\end{eqnarray}}
\newcommand{\bpf}{\begin{proof}}
\newcommand{\epf}{\end{proof}}
\newcommand{\ket}[1]{ | #1 \rangle}

\DeclareMathAlphabet{\mathpzc}{OT1}{pzc}{m}{it}
\newcommand{\Pv}{\mathcal{P}_V}

\newcounter{rowNo}
\definecolor{mygray}{gray}{0.5}

\begin{document}

%\title{Nonlocal strength in $(2, m, 2)$ and $(n, 2, 2)$ Bell scenarios  (???) }
\title{Strength and typicality of nonlocality in multisetting and multipartite Bell scenarios}

\author{Anna~de~Rosier}
\affiliation{Institute of Theoretical Physics and Astrophysics, Faculty of Mathematics, Physics and Informatics, University of Gda\'nsk, 80-308 Gda\'nsk, Poland}
\author{Jacek~Gruca}
\affiliation{Institute of Theoretical Physics and Astrophysics, Faculty of Mathematics, Physics and Informatics, University of Gda\'nsk, 80-308 Gda\'nsk, Poland}
\affiliation{International Centre for Theory of Quantum Technologies, University of Gda\'nsk, 80-308 Gda\'nsk, Poland}
\author{Fernando~Parisio}
\affiliation{Departamento de F\'isica, Federal University of Pernambuco, Recife, PE 50670-901, Brazil}
\author{Tam\'as~V\'ertesi}
\affiliation{Institute for Nuclear Research, Hungarian Academy of Sciences, H-4001 Debrecen, P.O. Box 51, Hungary}
\author{Wies{\l}aw~Laskowski}
\affiliation{Institute of Theoretical Physics and Astrophysics, Faculty of Mathematics, Physics and Informatics, University of Gda\'nsk, 80-308 Gda\'nsk, Poland}
\affiliation{International Centre for Theory of Quantum Technologies, University of Gda\'nsk, 80-308 Gda\'nsk, Poland}

%\date{\today}

\begin{abstract}
In this work we investigate the probability of violation of local realism under random measurements in parallel with the strength of these violations as described by resistance to white noise admixture. We address multisetting Bell scenarios involving up to 7 qubits.
As a result, in the first part of this manuscript we report statistical distributions of a quantity reciprocal to the critical visibility for various multipartite quantum states subjected to random measurements. The statistical relevance of different classes of multipartite tight Bell inequalities violated with random measurements is investigated. We also introduce the concept of typicality of quantum correlations for pure states as the probability to generate a nonlocal behaviour with both random state and measurement. Although this typicality is slightly above 5.3\% for the CHSH scenario, for a modest increase in the number of involved qubits it quickly surpasses 99.99\%.
\end{abstract}

\maketitle

\section{Introduction}

The probability of violation of local realism under random measurements, proposed in \cite{28procent}, 
has gained considerable attention as an operational measure of nonclassicality of quantum states \cite{Fernando15}. It has been demonstrated both numerically \cite{our1, our2, Barasinski} and analytically \cite{Fernando15,Lipinska18} that this quantity is a good candidate for a nonlocality measure. 
What is more, in \cite{Lipinska18} it was proved that this quantifier satisfies some natural properties and expectations for an operational measure of nonclassicality. 

In our approach to quantify nonlocality there are no prior assumptions about specific Bell inequalities. Instead we consider a joint probability distribution that is equivalent to the analysis of a full set of tight Bell inequalities in a given Bell scenario. The probability of violation is therefore defined as 
\be
\Pv(\rho)=\int f(\rho,\Omega) d\Omega,
\label{Pv}
\ee
where the integration variables correspond to all parameters that can vary within a Bell scenario and with
\be
f(\rho,\Omega)=\begin{cases}
    1, & \text{if settings lead to violations}\\
			 & \text{of local realism,}\\
    0, & \text{otherwise}.
  \end{cases}
\ee

Many quantum-informational protocols (as, for example, in quantum key distribution, secret sharing, and the reduction of communication complexity) require states that ``strongly'' violate Bell inequalities, however the concept of ``strength of violation'' is controversial in the literature.
Although definition \eqref{Pv} fairly captures the nonclassical extent of a state, it seems useful to put it together with another 
quantitative description which addresses the ``fragility'' of this nonclassicality against noise. Our approach enables us to report the probability 
that a state exhibits nonclassical correlations when random measurements are performed on it, and simultaneously measure the resistance to noise or to decoherence embodied by these quantum correlations \cite{Kaszlikowski00,Gruca10}. 
An extensive set of numerical results is presented and discussed in the next sections.

\section{Nonlocality strength distributions}

\subsection{Method}
Resistance to noise is understood as the amount of white noise admixture required to completely suppress the nonclassical character of the original correlations of a given state $\rho$. The state is now described by the following density operator: 
\be \rho(v) = v \rho + (1-v) \rho_{\textrm{whitenoise}}. 
\label{admixture}\ee 
The parameter $v$ is called the visibility of the state. 
For the states that reveal nonclassicality for a particular choice of observables, there always exists a critical visibility $v_{crit}$, such that for $v\leq v_{crit}$ a local realistic model can be constructed. The  critical visibility provides us with information about noise resistance of quantum correlations. The amount of noise for which the state becomes local will be called the strength of nonlocality $\mathcal{S} \equiv 1-v_{crit}$.

Determining the nonlocality strength is a linear programming problem. The set of linear equations, which have to be satisfied assuming local realism, is derived from linking the marginal quantum probabilities to preexisting joint probability distributions for all possible results of measurements performed on \eqref{admixture}. Obvious constraints such as normalization of probability and bounding the strength range ($0\leq \mathcal{S} \leq 1$) are also assumed.
An in-depth formulation of this problem and an explanation of the method harnessed to solve it can be found in \cite{Kaszlikowski00,Gruca10,Gruca14}.
 
The nonlocality strength is determined for a particular set of observables used and 
%is a function of angles parametrising the observables. 
we report its distribution obtained with a large statistics of random measurements applied to a given state. 
The measurement operators are sampled according to Haar measure, in the way described in \cite{our1,randomU}. 

In comparison with reference \cite{our1}, here we are interested not only in the summary probability that $\mathcal{S}>0$ for a random observable, but in a detailed probability distribution of achieving a specific value of the nonlocality strength $g(\mathcal{S})$.
We should also mention the approach in \cite{Gruca10}, where the main goal was to optimize $v_{crit}$ over all possible measurement settings to provide the minimal critical visibility for a given state $v_{crit}^{min}$ related here with the maximal nonlocality strength $\mathcal{S}^{max} = 1-v_{crit}^{min}$.

The probability of violation $\Pv$ does not provide any information about the strength of nonlocality, on the one hand. On the other hand, resistance against noise, although relevant, is not a proper nonlocality quantifier. A quantity that conveys both ingredients is likely to be useful. Therefore we define the average nonlocality strength:
\begin{equation}
\bar{\mathcal{S}} = \int_{0}^{\mathcal{S}^{max}}  \mathcal{S} ~  g(\mathcal{S}) ~ d\mathcal{S}. 
\end{equation}

\subsection{Results}

Our main results are collected in the form of histograms (Figs. \ref{hist2}-\ref{hist5}).
The  horizontal axis represents the nonlocality strength corresponding to the interval: $[\mathcal{S}-0.01, \mathcal{S})$ 
and on the vertical axis we have the probability density function (PDF).
All figures are normalized to a random set of settings drawn in the given experiment, so  the areas of the regions bounded by the plots are exactly the probabilities of violation, $\int g(\mathcal{S}) d\mathcal{S} = \mathcal{P}_V$.

The following observations can be drawn from the collected results.

\subsubsection{Two-qubit states}

We study two-qubit states of the form $\ket{\psi_{\rm GHZ}(\alpha)}=\cos\alpha\ket{00}+\sin\alpha\ket{11}$. Note that we can take the above Schmidt form without loss of generality, since the $\Pv$ function is invariant under local unitary transformations.

\begin{figure}[h!]
		\includegraphics[width=0.49\textwidth]{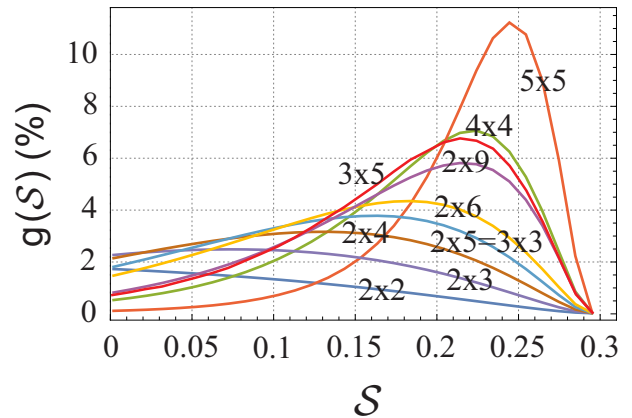}
	\caption{\label{hist2} Nonlocality strength distributions for the two-qubit GHZ state with various quantities of measurement settings for each qubit.}
\end{figure}

In Fig. \ref{hist2} we can observe the distribution characteristic's dependence on the number of measurement settings $m_1$ and $m_2$, 
respectively for the first and the second observer. When $m_1 \cdot m_2 < 10$ is satisfied, weaker violations are dominant and for $m_1 \cdot m_2 > 10$ this property is reversed. 

For $m_1 = m_2 \to \infty$ the plot should look like $\delta(\mathcal{S} - \mathcal{S}^{max})$ (compare with \cite{aaa}). Here $v_{crit}$ is related to $K_G(3)$, the Grothendieck constant of order three, as follows~\cite{agt}: 
$K_G(3)=1/v_{crit}=1/(1-\mathcal{S}^{max})$. From the known best upper~\cite{hirsch} and lower~\cite{dbv} bounds to $K_G(3)$, $\mathcal{S}^{max}$ is bounded as $0.3036\le\mathcal{S}^{max}\le 0.3171$. With infinitely many settings, one can always find such settings for $\mathcal{S}<\mathcal{S}^{max}$ that local realism is violated. We can also notice the coincidence of nonlocality strength distributions for the $2\times 5$ and $3\times 3$ measurement settings (these are the ones with the product closest to each other). It is noted that when $\rho$ in Eq.~(\ref{admixture}) is the two-qubit maximally entangled state (i.e.~$\alpha=45^{\circ}$) the single-party expectation values of $\rho(v)$ vanish. In this case our analysis can be restricted to the Bell polytope involving only joint correlation terms. In this reduced space, the polytope is often called the correlation polytope, and the only facets in the $3\times m$ scenarios for $m\ge 2$ are the variants of the CHSH inequality~\cite{garg}. 
Hence, the similarity of the curves corresponding to $2\times 5$ and $3\times 3$ scenarios has to relate to statistical considerations: Applying more than two settings for at least one party increases the chance of violation of one of the CHSH-Bell inequalities simply due to statistical reasons. We observe a similar behavior  for $3\times 5$ and $4 \times 4$ cases. 

\begin{figure}[h!]
	\includegraphics[width=0.49\textwidth]{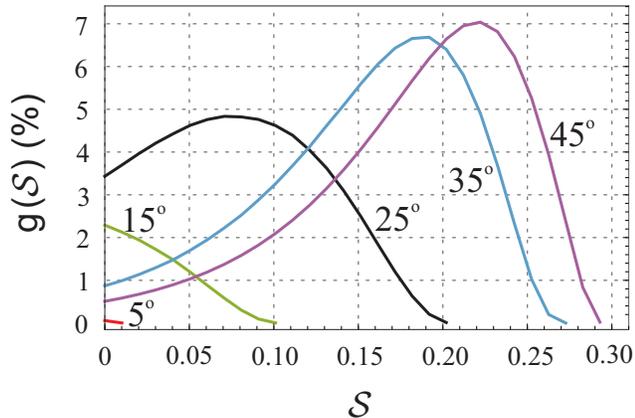}
	\caption{\label{hist2deg} Nonlocality strength distributions for the two-qubit generalized GHZ($\alpha$) state with $4\times 4$ measurement settings and with selected values of $\alpha$.}
\end{figure}

From Fig.~\ref{hist2deg} we know that the more symmetric the state is the higher are the dominant nonlocality strengths.

The nonlocality strength for all considered numbers of settings for $\alpha=45^{\circ}$ are constant and equal to $1-1/\sqrt{2}$ (this value of $\mathcal{S}^{max}$, corresponding to $v_{crit}=1/\sqrt 2$ for the scenarios $m\times m$ with $m\le 5$, comes from the studies in Refs.~\cite{ds,dip}). However, the averaged nonlocality strengths decrease/increase with the number of settings and they are equal to: 0.028 for $2\times2$, 0.110 for $3\times3$, 0.178 for $4\times4$ and 0.218 for $5\times5$ scenario.

\subsubsection{Four-qubit states}

\begin{figure}[!h]
	\includegraphics[width=0.49\textwidth]{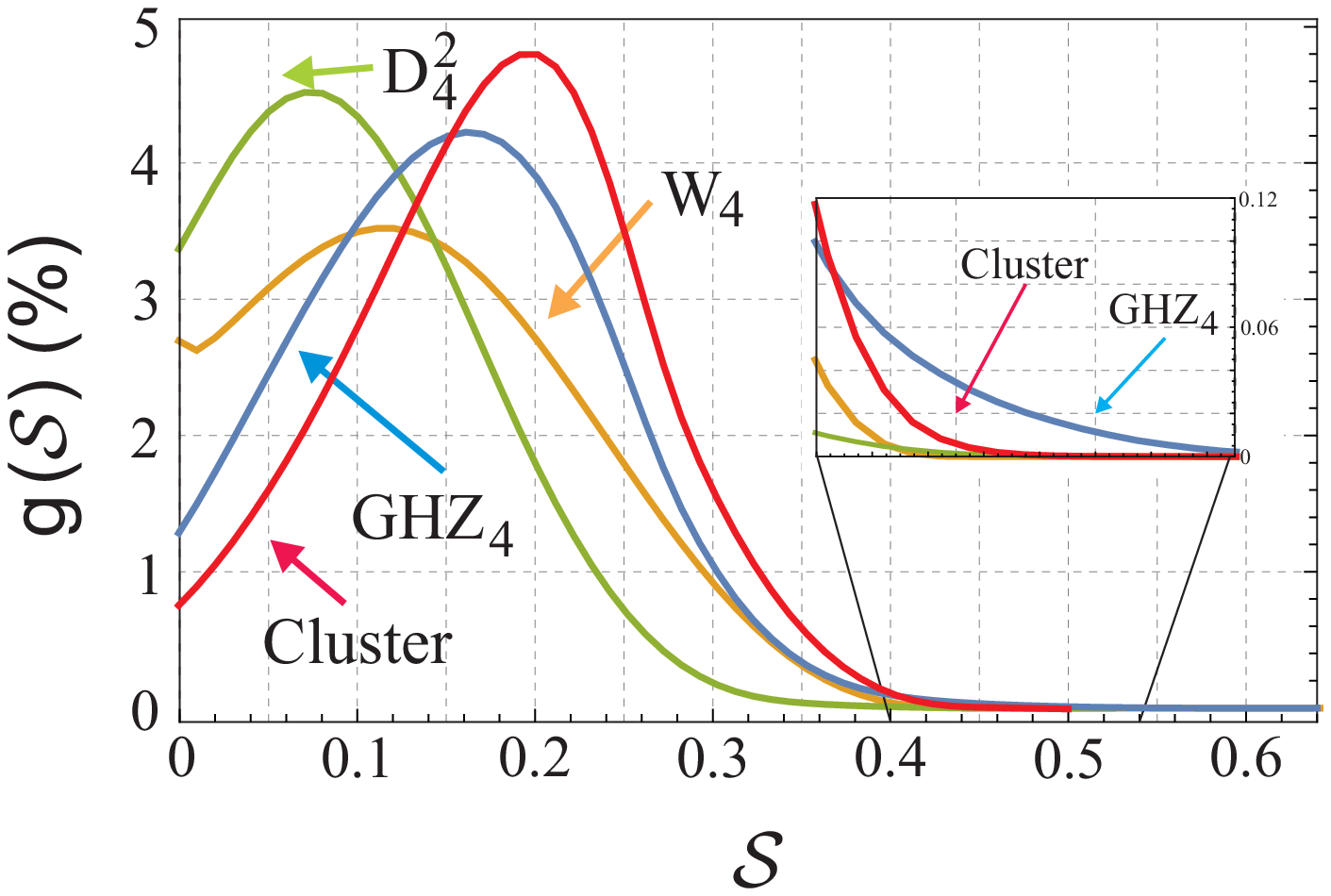}
	\caption{\label{hist4} Nonlocality strength distributions for four-qubit states.}
\end{figure}

Very strong violations (high nonlocality strength) are more probable for the GHZ state than for other prominent families of four-qubit states (Dicke, W, Cluster). For instance, for nonlocality strength 0.45, the GHZ state violates Bell inequalities 7.4 times more likely than the Cluster state. However, the Cluster state surpasses the GHZ state in intermediate values of nonlocality strength (see Fig. \ref{hist4}), where for example for $\mathcal{S}=0.35$ the violations are observed 1.7 times more often. This suffices to make the Cluster states attain highest probabilities of violation among all considered states. Also the averaged nonlocality strength for the Cluster state (0.1843) is higher than for the GHZ state (0.1624). Here we considered two settings per party.

\begin{figure}[!h]
	\includegraphics[width=0.49\textwidth]{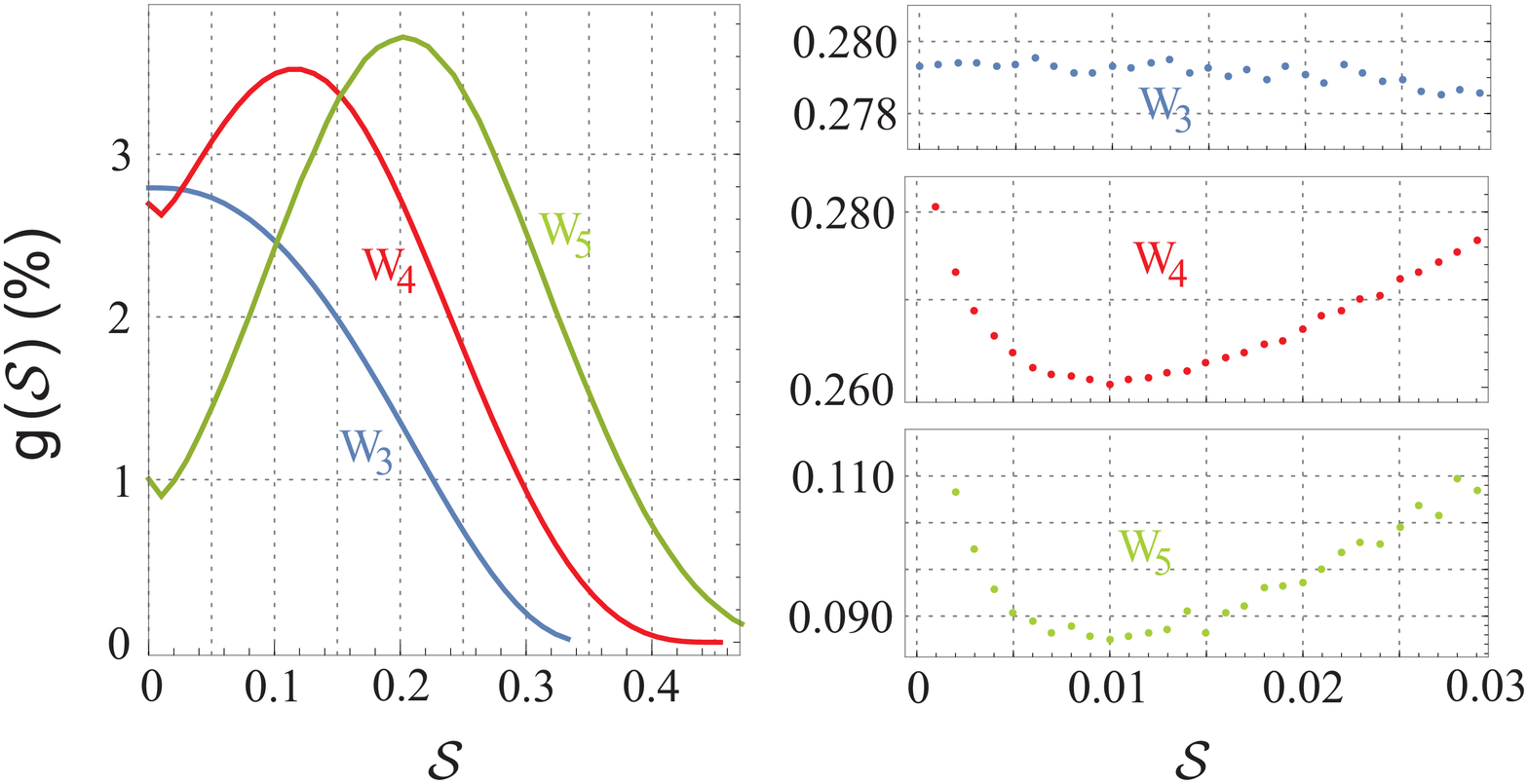}
	\caption{\label{histW} Unexpected behavior of nonlocality strength distributions close to $\mathcal{S}=0$ for the W states.}
\end{figure}

When analyzing the histogram for the four-qubit W state we noticed a surprising behavior, namely, a dip for strengths close to 0.02 (see Fig.\ref{histW}). This was also observed for the five-qubit W state, but not for three-qubits. 
A possible explanation of this feature could be the
fact that there are more than one relevant Bell inequality for the
considered cases with different functions representing the nonlocality strength.
The total nonlocality strength  is a combination of the
strengths for those particular inequalities, which may result in several extremes.

\subsubsection{Five-qubit states}

\begin{figure}[h!]
	\includegraphics[width=0.49\textwidth]{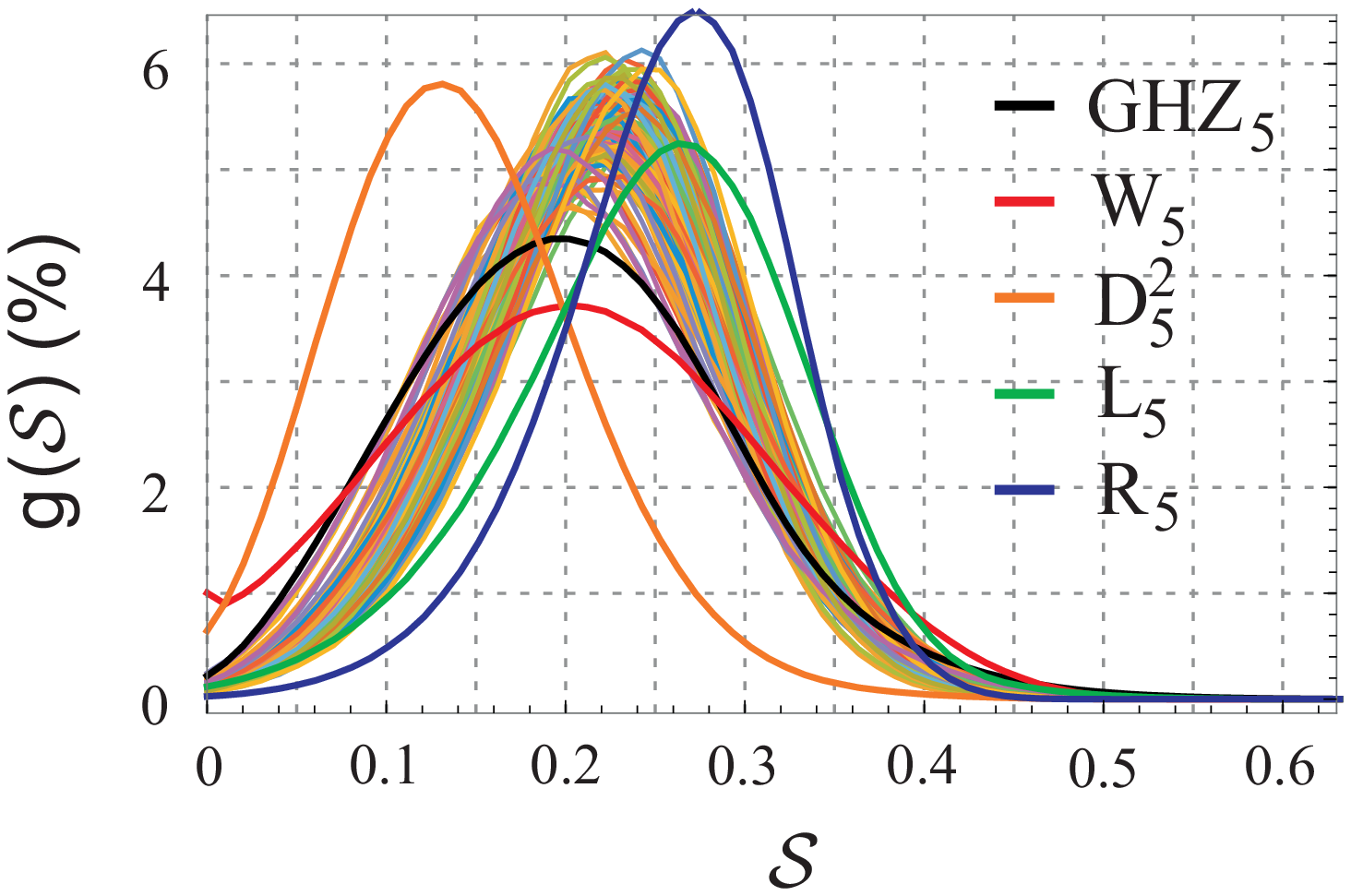}
	\caption{\label{hist5} Nonlocality strength distributions for especial and random five qubit states.}
\end{figure}

In Fig. \ref{hist5} we can compare especial five-qubit states (GHZ, Dicke, W, linear- and ring cluster states \cite{GRAPH}) with 100 random pure states. These states are distinguishable either because they have maxima at very different values $(|R_5\rangle, |L_5\rangle,|D_5^2\rangle)$ or because they present larger variances ($|GHZ\rangle, |W\rangle$).

\section{Statistical relevance of facet inequalities}

We observe that the probability of violation rapidly increases with the number of measurement settings.

The first explanation of that fact is statistical. By increasing the number of settings, we increase the probability that some of them violate the Bell inequality involving only two settings (CHSH). For instance, for the experiment with the two qubit GHZ state and 5 measurement settings per party one can identify the following families of Bell inequalities:
\be
\langle a_1 b_1 + a_1 b_2 + a_2 b_1 - a_2 b_2\rangle \leq 2, 
\label{i1}
\ee
\be
\langle a_1 b_1 - a_3 b_1 + a_4 b_1 + a_5 b_1 - a_3 b_2 - a_4 b_2 - a_1 b_4  \nonumber \\
+ a_3 b_4 - a_4 b_4 + a_5 b_4 – 2 a_1 b_5 - a_3 b_5 + a_4 b_5\rangle \leq 6, \label{i2}
\ee
\be
\langle a_1 b_1 + a_2 b_1 - a_4 b_1 - a_5 b_1 - a_1 b_3 + a_2 b_3 - a_4 b_3 \nonumber  \\
+ a_5 b_3 - a_1 b_4 + a_2 b_4 - 2 a_3 b_4 + a_4 b_4 - a_5 b_4 \label{i3}  \\
- a_1 b_5+ a_2 b_5 + 2 a_3 b_5 + a_4 b_5 - a_5 b_5\rangle \leq 8, \nonumber
\ee
\be
\langle - a_1 b_1 - a_2 b_1 - a_3 b_1 + 2 a_4 b_1 +  a_5 b_1 + a_1 b_2 \nonumber \\
 + a_4 b_2 - a_5 b_2 + a_1 b_3 - a_3 b_3 + a_4 b_3 - a_5 b_3    \nonumber \\
 - a_3 b_4 + a_4 b_4 + 2 a_5 b_4 + 2 a_3 b_5 + a_4 b_5 + a_5 b_5  \label{i4}\\  
+ a_2 b_4 + a_3 b_2 + a_1 b_4 \rangle\leq 10. \nonumber
\ee
Here $\langle a_i b_j \rangle$ denotes an expectation value of the correlation measurement in which the first and the second observers measure observables $a_i$ and $b_j$, respectively. Each family contains many equivalent inequalities. For example, the first family (\ref{i1}) is obtained by replacing settings $a_i \to \pm a_k$, and $b_j \to \pm b_l$, where $k,l \in \{1,2,3,4,5\}$. 

We note that all the above Bell inequalities are tight, that is, they define facets of the Bell local polytope (and they are facets of the correlation polytope too). The inequality~(\ref{i2}) is a genuine $(4\times 5)$-setting Bell inequality, whereas (\ref{i3}) and (\ref{i4}) are genuine $(5\times 5)$-setting Bell inequalities.

The highest violation strength is observed for inequalities belonging to family (\ref{i1}) in 99.1\% of random sets of settings, (\ref{i2}) in 0.85\%, (\ref{i3}) in 0.04\% and (\ref{i4}) in 0.01\%. This means that in almost all cases we effectively use only two out of five settings. 

A distinct effect is observed in the case of the three qubit W state. 
In this case 14\% of observed violations really involve three measurement settings. 
One of the examples of a genuine $3\times3\times3$ inequality is:
\be
&\langle a_1 + 5 a_2 - 5 a_3 +  b_1 -  a_1 b_1 -  a_2 b_1 + 3 a_3 b_1 +  3 b_2 +  a_1 b_2 &\nonumber \\ &- 2 a_3 b_2  +  b_3 +  a_1 b_3 +  c_1 - 2 a_1 c_1 -  2 a_2 c_1 +  a_3 c_1 + 2 b_1 c_1& \nonumber \\& + 4 a_1 b_1 c_1 
-  a_2 b_1 c_1 + a_3 b_1 c_1 - 2 b_2 c_1 + 2 a_1 b_2 c_1 + 3 a_2 b_2 c_1&\nonumber \\  &- 3 a_3 b_2 c_1 - 5 b_3 c_1 + 4 a_2 b_3 c_1 - 3 a_3 b_3 c_1+  a_2 c_2 +  a_3 c_2&\nonumber \\ & + a_1 b_1 c_2 - 3 a_2 b_1 c_2 - 2 a_3 b_1 c_2 - 3 b_2 c_2 - 3 a_3 b_2 c_2  + 3 b_3 c_2& \nonumber \\ &+  a_1 b_3 c_2 - 4 a_2 b_3 c_2 + 2 c_3+  a_1 c_3- 2 a_2 c_3 + a_3 c_3- 3 b_1 c_3 & \nonumber \\ &- 2 a_1 b_1 c_3 +  a_2 b_1 c_3 - 4 a_3 b_1 c_3 + 
2 b_2 c_3 - 3 a_1 b_2 c_3 &\nonumber \\& + 3 a_2 b_2 c_3 + 2 a_3 b_2 c_3 - 3 b_3 c_3 - 3 a_3 b_3 c_3 \rangle\leq 23.&
\ee

\section{Typicality of nonlocality}

It is known that the set of multipartite entangled states is large \cite{SZAREK}. Obviously almost all random pure states are entangled. Now one can ask a similar question concerning a more demanding property -- nonlocality. What is the typical probability of violation $T_{V}$ for a randomly sampled pure state? In this problem we specify only the number of qubits $N$. For two observers we analyze experiments up to 11 settings per side. For three and more observers, we employed Bell scenarios involving only two measurement settings. For a given state we verify the violation only for a single randomly chosen set of settings.

The results are presented in Tab. \ref{tab-pure}. Already for $N=4$ we observe a quite high typical probability of violation ($T_{V} >93\%$). For $N>4$ it is practically equal to $100\%$. This means that almost all states violate local realism for any settings. The result seems to be stronger than in the case of ``typicality of entanglement'', where measurements are optimized. Also, the typical nonlocality strength $ T_{\mathcal{S}} $ increases with the number of parties and settings. The value of $T_{\mathcal{S}}$ for two qubits and infinitely many measurement settings can be bounded from below by 0.1436. This value comes from the assumption that all facets of the corresponding Bell polytope are defined only by the variants of the CHSH inequalities. The nonlocality strength can be calculated using Horodecki’s formula \cite{Hor} and averaging over all pure random states.

\begin{center}
\begin{table}
\begin{tabular}{c c c c c}
\hline \hline
$N$ & Settings & Statistics & $T_V(\%)$ & $ T_{\mathcal{S}} $\\ 
\hline
2 & $2\times2$ &  $10^9$ & 5.32 &0.004\\ 
  & $3\times3$ &  $10^9$ & 21.99 &0.019\\ 
  & $4\times4$ &  $10^8$ & 38.43 &0.038\\ 
  & $5\times5$ &  $10^7$ & 50.04 &0.054\\ 
  & $6\times6$ &  $10^7$ & 57.98 &0.068\\ 
  & $7\times7$ &  $10^6$ & 63.63 &0.079\\ 
  & $8\times8$ & $10^5$ & 67.83  &0.087\\ 
  & $9\times9$ & $10^4$ & 71.23  &0.093 \\ 
  & $10\times10$ & $10^4$ & 74.34&0.097 \\ 
  & $11\times11$ & $10^3$ & 76.80 &0.101\\ \hline
3 & $2\times2\times2$ &   $10^7$ & 42.96& 0.034\\ 
4 & $2\times2\times2\times2$ & $10^8$ & 93.28& 0.123\\ 
5 & $2\times2\times2\times2\times2$  & $10^7$ & 99.88 &0.222\\ 
6 & $2\times2\times2\times2\times2\times2$  & $10^6$ & $>99.99$ &0.306\\ 
7 & $2\times2\times2\times2\times2\times2\times2$  & $10^4$ & $>99.99$&0.377 \\ 
\hline
\hline
\end{tabular}
\caption{\label{tab-pure} Typical probability of violation $T_{V}$ and typical nonlocality strength  $T_{\mathcal{S}}$ for pure random qubit states and random measurements (one random measurement per random state).}
\end{table}
\end{center}

\section{Closing remarks}

In this paper we employed linear programming as a useful
tool to analyze the nonclassical properties of quantum states.
We introduced nonlocality strength as a resistance to white noise admixture and verify its statistical properies.
Most of the conclusions were presented in the previous sections. Here we want to stress
that the overall message of the obtained results is that nonlocality is a typical phenomenon for multipartite states, i.e. the probability that a random multipartite state violates some Bell inequaltiy for a random set of measurement settings is close to one.

\section{Acknowledgements}

We thank Lukas Knips and Harald Weinfurter for valuable discussions. 
WL acknowledges the support by DFG (Germany) and NCN (Poland) within the joint funding initiative ``Beethoven2'' (2016/23/G/ST2/04273). JG and WL acknowledge partial support by the Foundation for Polish Science (IRAP project, ICTQT, contract no. 2018/MAB/5, co-financed by EU via Smart Growth Operational Programme)." FP thanks financial support from Coordena\c{c}\~ao de Aperfei\c{c}oamento de Pessoal de N\'{\i}vel Superior (CAPES), Funda\c{c}\~ao de Amparo \`a Ci\^encia e Tecnologia do Estado de Pernambuco (FACEPE), and Conselho Nacional de Desenvolvimento Cient\'{\i}fico  e Tecnol\'ogico (CNPq). TV was supported by the National Research, Development and Innovation Office NKFIH (Grant No. KH125096).

\end{document}